\pdfoutput=1
\documentclass[sigconf,screen]{acmart}

\newcommand{\argmax}{\mathop{\mathrm{argmax\,}}}

\newcommand{\boldA}{{\boldsymbol{A}}}

\newcommand{\boldI}{{\boldsymbol{I}}}

\newcommand{\boldb}{{\boldsymbol{b}}}

\newcommand{\boldv}{{\boldsymbol{v}}}

\usepackage{bbm}
\usepackage{mathtools}
\usepackage[ruled, algo2e, vlined]{algorithm2e}
\usepackage{tcolorbox}
\usepackage{xcolor}

\SetCommentSty{mycommfont}

\usepackage{CJKutf8}

\allowdisplaybreaks

\AtBeginDocument{%
  \providecommand\BibTeX{{%
    \normalfont B\kern-0.5em{\scshape i\kern-0.25em b}\kern-0.8em\TeX}}}

\copyrightyear{2022}
\acmYear{2022}
\setcopyright{rightsretained}
\acmConference[WSDM '22]{Proceedings of the Fifteenth ACM International Conference on Web Search and Data Mining}{February 21--25, 2022}{Tempe, AZ, USA}
\acmBooktitle{Proceedings of the Fifteenth ACM International Conference on Web Search and Data Mining (WSDM '22), February 21--25, 2022, Tempe, AZ, USA}
\acmDOI{10.1145/3488560.3498462}
\acmISBN{978-1-4503-9132-0/22/02}

\begin{document}

\title{Retrieving Black-box Optimal Images from External Databases}

\author{Ryoma Sato}
\email{r.sato@ml.ist.i.kyoto-u.ac.jp}
\affiliation{%
  \institution{Kyoto University / RIKEN AIP}
}


\begin{abstract}
Suppose we have a black-box function (e.g., deep neural network) that takes an image as input and outputs a value that indicates preference. How can we retrieve optimal images with respect to this function from an external database on the Internet? Standard retrieval problems in the literature (e.g., item recommendations) assume that an algorithm has full access to the set of items. In other words, such algorithms are designed for service providers. In this paper, we consider the retrieval problem under different assumptions. Specifically, we consider how users with limited access to an image database can retrieve images using their own black-box functions. This formulation enables a flexible and finer-grained image search defined by each user. We assume the user can access the database through a search query with tight API limits. Therefore, a user needs to efficiently retrieve optimal images in terms of the number of queries. We propose an efficient retrieval algorithm \textsc{Tiara} for this problem. In the experiments, we confirm that our proposed method performs better than several baselines under various settings.
\end{abstract}

\ccsdesc[500]{Information systems~Web searching and information discovery}
\ccsdesc[500]{Information systems~Users and interactive retrieval}

\keywords{Information Retrieval, Web Searching, Linear Bandits, Private Recommender Systems}

\maketitle

\begin{table*}[tb]
    \centering
    \caption{Symbols, Definitions, and Examples.}
    \vspace{-0.15in}
    \scalebox{0.9}{
    \begin{tabular}{lll} \toprule
        \textbf{Symbol} & \textbf{Definition} & \textbf{Example} \\ \midrule
        $\mathcal{X}$ & The set of images in the image database & The set of all images uploaded on Flickr \\
        $\mathcal{T}$ & The set of tags & The set of all tags on Flickr \\
        $f\colon \mathcal{X} \to \mathbb{R}$ & The objective black-box function & A deep neural network that computes a preference score of an image \\
        $\mathcal{O}\colon \mathcal{T} \to \mathcal{X} \times 2^{\mathcal{T}}$ & The oracle for image search & A wrapper of \texttt{flickr.photos.search} API \\
        $\mathcal{T}_\text{ini} \subseteq \mathcal{T}$ & Initially known tags & $100$ popular tags obtained by \texttt{flickr.tags.getHotList} API \\
        $B \in \mathbb{Z}_+$ & A budget of oracle calls & $B = 500$ \\ \bottomrule
    \end{tabular}
    }
    \label{tab: symbol}
    \vspace{-0.15in}
\end{table*}

\section{Introduction}

As the amount of information on the Internet continues to drastically increase, information retrieval algorithms are playing more important roles. In a typical situation, a user of a system issues a query by specifying keywords, and an information retrieval algorithm retrieves the optimal items with respect to the query words. Here, the retrieval algorithm is designed by the service provider, not by the user. The uses of information systems have become divergent, and various retrieval algorithms have therefore been proposed, e.g., a cross-modal image search \cite{cao2016deep, kordan2018deep}, complex query retrieval \cite{nie2012harvesting}, and conversational recommender systems \citep{sun2018conversational, christakopoulou2016towards}.

Deep neural networks have achieved state-of-the-art performances in computer vision tasks \cite{krizhevsky2012imagenet, he2016deep}, notably image retrieval \cite{babenko2015aggregating, gordo2016deep, bell2015learning, niu2018neural}. In a conventional setting of image retrieval, algorithms assume that they have full access to the image database. A straightforward method under this setting is to evaluate all images in the database and return the one with the maximum score. When the image database is extremely large, two-step methods are used, i.e., a handful of images are retrieved through fast retrieval algorithms such as a nearest neighbor search, and the results are ranked using sophisticated algorithms. Hash coding further improves the effectiveness and efficiency \cite{lai2015simultaneous, liu2016deep}.

In this study, we consider information retrieval under a different scenario. Whereas most existing studies have focused on how a service provider can improve the search algorithms, we focus on how a user of a service can effectively exploit the search results. Specifically, we consider a user of a service builds their own scoring function. The examples of the scoring functions are as follows.

\noindent \textbf{Example 1 - Favorite Image Retrieval:} A user trains deep neural networks using a collection of favorite images found in different services and wants to retrieve images with similar properties in a new service.

\noindent \textbf{Example 2 - Similar Image Retrieval:} Deep convolutional neural networks are known to have a superior ability to extract useful image features \cite{babenko2015aggregating, gatys2016image, bell2015learning}. Although some online services provide similar image search engines, users do not have full control of the search. For example, even if the service provides a similar texture-based image search engine, some users may want to retrieve similar images based on the semantics. The user-defined score function allows image searches at a finer granularity.

\noindent \textbf{Example 3 - Fair Image Retrieval:} A search engine can be unfair to some protected attributes. For example, when we search for images through a query ``president,'' an image search engine may retrieve only male president images \cite{singh2018fairness}. Some users may want to use their own scoring function, e.g., scoring male and female images equally.

Recent advancements in deep learning, such as self-supervised contrastive learning \cite{chen2020simple, he2020momentum} and meta-learning \cite{finn2017model}, enable the training of deep neural networks with a few labeled samples. In addition, many pre-trained models for various vision tasks have been released on the Internet. Machine learning as a service platform, such as Microsoft Azure Machine Learning Studio and AWS Machine Learning, has also made it easy to build a machine learning model. These techniques and services enable an individual to easily build their own black-box scoring function. Suppose a user has already built a black-box function she wants to optimize. How can she retrieve optimal images with respect to this function from an image database on the Internet? 

Under this setting, an individual cannot evaluate all images in the image database because it contains a significant number of images. It is also impossible for an individual to build a search index (e.g., a hash index) because of both limited access to the database and the insufficient computational resources of the individual. These limitations prohibit the use of standard image retrieval algorithms.

In this paper, we assume that a user can access the database through a search query alone and that a tight query budget exists. In addition, we assume little knowledge about the database to apply our method to a new environment. We formulate the problem through the lens of the multi-armed bandit problem and propose a query efficient algorithm, \textsc{Tiara}, with the aid of pre-trained representative word embeddings. We confirm the effectiveness of the proposed method under various settings, including the online Flickr environment.

It is noteworthy that Bayesian optimization and optimization on deep neural networks \cite{erhan2009visualizing, simonyan2014deep, nguyen2016synthesizing} also aim to optimize black-box or complex functions. However, these methods assume a continuous and simple optimization domain, typically the entire Euclidean space $\mathbb{R}^d$, a hypercube $[0, 1]^d$, or a unit ball $\{x \mid \|x\| \le 1\}$. By contrast, we aim to \emph{retrieve} optimal images from the \emph{fixed database}. Therefore, the optimization domain is a discrete set of images. Although there are several methods for discrete black-box optimization \cite{baptista2018baysian, oh2019combinatorial}, they also assume that the optimization domain is simple, e.g., $\{0, 1\}^d$, or at least they have full access to the optimization domain. Under our setting, an algorithm does not even know the entire optimization domain, i.e., the image database. This limitation causes numerous challenges, as we describe in the following sections.

The contributions of this paper are summarized as follows:

\begin{itemize}
    \item We formulate the black-box optimal image retrieval problem. This problem examines how a user of an online service can effectively exploit a search engine, whereas most existing studies focus on how a service provider can improve a search engine.
    \item We propose \textsc{Tiara}, an effective method for this problem for the first time. \textsc{Tiara} is a general algorithm that works in various situations and retrieves optimal black-box images with few API queries.
    \item We investigate the effectiveness of \textsc{Tiara} using many real-world data. Notably, we conduct ``in the wild'' experiments on the real-world Flickr environment and confirm that \textsc{Tiara} can be readily used in real-world applications.
\end{itemize}

\noindent \textbf{Reproducibility:} Our code is available at \url{https://github.com/joisino/tiara}.

\section{Background}

\subsection{Notations}

Let $\mathcal{X}$ be a set of images in the image database. We do not know the exact set of $\mathcal{X}$. Let $f: \mathcal{X} \to \mathbb{R}$ be a black-box function that evaluates the value of an image. For example, $f$ measures the preference of the user in favorite image retrieval and measures the similarity in similar image retrieval. The notations are summarized in Table \ref{tab: symbol}.

\subsection{Problem Formulation}

In this section, we formulate the problem of black-box optimal image retrieval. We observe that many image databases, such as Flickr and Instagram, support (hash) tag-based search. We assume that we can search for images by specifying a tag through an API. For example, in the Flickr case, we use the \texttt{flickr.photos.search} API.

For a formal discussion, we generalize the function of tag search APIs. Let $\mathcal{T} \subset \Sigma^*$ be the set of tags, where $\Sigma^*$ is the set of strings. We formalize a tag search API as a (randomized) oracle $\mathcal{O}$ that takes a tag as input and returns an image with the tag. We assume that $\mathcal{O}$ always returns different images even when we query the same tag twice. In addition to the image itself, we assume that the oracle $\mathcal{O}$ returns the set of tags of the image. Therefore, for any tag $t \in \mathcal{T}$, $\mathcal{O}(t) \in \mathcal{X} \times 2^\mathcal{T}$. Let $\mathcal{O}(t).\texttt{image} \in \mathcal{X}$ denote the returned image and $\mathcal{O}(t).\texttt{tags} \in 2^\mathcal{T}$ denote the returned tags. As the oracle returns an image with the query tag, $t \in \mathcal{O}(t).\texttt{tags}$ always holds.

We find that myriad tags exist in real-world services, and we cannot know the entire tag set. Therefore, we assume that we know only a fraction $\mathcal{T}_\text{ini}$ of the tag set. Here, $\mathcal{T}_\text{ini}$ can be constructed by browsing the online service or retrieving popular tags through an API, e.g., \texttt{flickr.tags.getHotList}, in the Flickr example.

We assume that there is a budget $B \in \mathbb{Z}_+$ for the oracle call. For example, in the Flickr case, There is an API rate limit of $3600$ calls per hour. Thus, it is natural to assume that we can use at most $3600$ API calls in one task. If we retrieve many images within a short period of time, the API limits become tighter for each retrieval. To summarize, the black-box optimal image retrieval problem is formalized as follows.

\begin{tcolorbox}[colframe=gray!20,colback=gray!20,sharp corners]
\textbf{Black-box optimal image retrieval.}

\textbf{Given:}
\begin{itemize}
    \item Black box function $f\colon \mathcal{X} \to \mathbb{R}$
    \item Oracle $\mathcal{O}\colon \mathcal{T} \to \mathcal{X} \times 2^{\mathcal{T}}$
    \item Known tags $\mathcal{T}_\text{ini} \subseteq \mathcal{T}$
    \item Budget $B \in \mathbb{Z}_+$
\end{itemize}

\textbf{Goal:} Find an image $x \in \mathcal{X}$ in the image database with as high an $f(x)$ as possible within $B$ accesses to the oracle.
\end{tcolorbox}

\section{Proposed Method}

In this section, We introduce our proposed method, \textsc{Tiara} (\underline{T}ag-based \underline{i}m\underline{a}ge \underline{r}etriev\underline{a}l).

\subsection{Bandit Formulation} \label{sec: bandit}

We first propose regarding the black-box optimal image retrieval problem as a multi-armed bandit problem \cite{lattimore2020bandit, slivkins2019introduction}. Specifically, we regard a tag as an arm, the budget $B$ as the time horizon, and the objective value $f(\mathcal{O}(t).\texttt{image})$ as the reward when we choose arm $t$. This formulation enables us to use off-the-shelf multi-armed bandit algorithms, such as UCB \cite{auer2002finite}, $\varepsilon$-greedy, and Thompson sampling \cite{thompson1933likelihood}. This formulation is the basics of our proposed algorithm. However, there are several challenges to a black-box optimal image retrieval problem. First, we do not initially know all arms but only a fraction $\mathcal{T}_\text{ini}$ of arms. Considering $\mathcal{T}_\text{ini}$ alone leads to suboptimal results because $\mathcal{T}_\text{ini}$ does not contain the best arm in general. Second, myriad arms exist, and the number $|\mathcal{T}|$ of arms is larger than budget $B$ in practice. Standard multi-armed bandit algorithms first explore all arms once. However, they are unable to even finish this initial exploration phase under a tight budget constraint.

The first problem is relatively easy to solve. We obtain tags $\mathcal{O}(t).\texttt{tags}$ when we choose arm $t$. These tags may contain new tags. We can add such tags into the known tag set and gradually grow the known set. A good bandit algorithm will choose relevant tags, and the returned tags will contain relevant tags. For example, suppose that the black-box function $f$ prefers cat images. A good bandit algorithm will choose ``cat'' and ``animal'' tags and obtain cat images accompanied by many cat-related tags. Even if the budget is so tight that we cannot know all tags within the budget, a good bandit algorithm ignores irrelevant tags and prioritizes the collection of many relevant tags. The second problem is essential and difficult to solve. We investigate how to improve the query efficiency using tag embeddings in the following sections.

\vspace{0.1in}
\noindent \textbf{Discussion on the max-bandit problem.} It should be noted that standard bandit algorithms aim to maximize a cumulative reward or regret, whereas the black-box optimal image retrieval problem aims to find an image with the maximum value. These goals are not exactly the same. The problem of obtaining a maximum reward is known as a max $K$-armed bandit \cite{cicirello2005max} or extreme bandit \cite{carpentier2014extreme}. However, we adhere to the standard bandit setting in this study for the following reasons: First, we assume that the reward distribution is Gaussian-like, whereas existing max-$K$-armed bandit algorithms \cite{cicirello2005max, carpentier2014extreme} mainly assume that rewards are drawn from extreme value distributions, such as the GEV distribution. We find that this is not the case in the applications considered herein, i.e., image retrieval. Second, under our setting, an arm with a high expected value often leads to a high maximum value because such an arm usually represents the concept captured by the black-box function $f$. By contrast, max-bandit algorithms focus on detecting heavy tail arms with possibly low average rewards. We find that this is not necessary for our setting. Finally, we experimentally confirm that our proposed method already shows a superior empirical performance even though it does not directly maximize the maximum objective value. We leave the exploration of the max-bandit algorithms in our setting for future work.

\subsection{Tag Embedding} \label{sec: embedding}

\setlength{\textfloatsep}{5pt}
\begin{algorithm2e}[t]
\caption{\textsc{Tiara}}
\label{algo}
\DontPrintSemicolon 
\nl\KwData{Known tags $\mathcal{T}_\text{ini} \subseteq \mathcal{T}$, Black box function $f\colon \mathcal{X} \to \mathbb{R}$, Budget $B \in \mathbb{Z}_+$, Oracle $\mathcal{O}\colon \mathcal{T} \to \mathcal{X} \times 2^{\mathcal{T}}$, Tag Embedding $\{\boldv_t \in \mathbb{R}^d \mid t \in \mathcal{T}\}$, Regularization coefficient $\lambda \in \mathbb{R}_+$, Exploration coefficient $\alpha \in \mathbb{R}_+$.}
\nl\KwResult{An image $x \in \mathcal{X}$ in the image database with as high $f(x)$ as possible.}
    \nl $\boldA \leftarrow \lambda \boldI_{d \times d}$ \tcp*{Initialize $\boldA$}
    \nl $\boldb \leftarrow \mathbf{0}_{d}$  \tcp*{Initialize $\boldb$}
    \nl $\mathcal{T}_\text{known} \leftarrow \mathcal{T}_\text{ini}$  \tcp*{Initialize with the initial tags}
    \nl \For{$i \gets 1$ \KwTo $B$}{
    \nl     $s_t \leftarrow \boldv_t^\top \boldA^{-1} \boldb + \alpha \sqrt{\boldv_t^\top \boldA^{-1} \boldv_t} \quad \forall t \in \mathcal{T}_\text{known}$ \tcp*{Scores}
    \nl     $t_i \leftarrow \argmax_{t \in \mathcal{T}_\text{known}} s_t$ \tcp*{Choose the best arm}
    \nl     $r_i \leftarrow \mathcal{O}(t_i)$ \tcp*{Issue an query}
    \nl     \For{$t \in r_i.\texttt{tags}$}{
    \nl         $\boldA \leftarrow \boldA + \boldv_t \boldv_t^\top$ \tcp*{Update $\boldA$}
    \nl         $\boldb \leftarrow \boldb + f(r_i.\texttt{image}) \cdot \boldv_t$ \tcp*{Update $\boldb$}
    \nl         $\mathcal{T}_\text{known} \leftarrow \mathcal{T}_\text{known} \cup \{t\}$ \tcp*{Insert $t$ to $\mathcal{T}_\text{known}$}
            }
        }
    \nl \textbf{return} $\argmax_{x \in \{r_i.\texttt{image} \mid i = 1, \cdots, B\}} f(x)$ \tcp*{Best image}
\end{algorithm2e}

Because we cannot choose each arm even once on average, we need to estimate the value of each arm without observing the reward from it. We estimate the value of one arm from the rewards obtained from the other arms. The key is how to define the similarities of the arms. As the challenge here, we assume a new image database, and owing to the tight API limit, it is difficult to learn the similarities from the current environment in an online manner. To tackle this problem, we utilize external resources. We find that a tag in an image database is usually described through natural language and typically is a word or composition of words. We use a pre-trained word embedding \cite{mikolov2013efficient, pennington2014glove} to define the similarities between tags. Specifically, we first decompose a tag into a bag of words by non-alphabetical symbols, e.g., a white space.
We use the mean of the word embeddings in the bag of words as the tag embedding.

Owing to the tag embeddings, we can infer the value of an arm from similar arms. For example, if the reward from a ``cat'' tag is high (resp. low), we can assume the black-box function is highly (resp. rarely) related to cats, and we can infer that similar tags with similar embeddings, such as ``Aegean cat'' and ``animal,'' are also valuable (resp. irrelevant) without actually querying them.

\subsection{Tiara}

Our proposed method combines the aforementioned tag embeddings with a bandit algorithm. Specifically, we utilize LinUCB \cite{li2010contextual}. Because there are no contextual features, we use only features of arms. We define the feature of an arm as the tag embedding introduced in Section \ref{sec: embedding} and apply LinUCB to this feature. We call this variant \textsc{Tiara}-S, where S stands for ``single'' and ``simple.'' However, we found that there are too many tags, and LinUCB is still inefficient for learning a reward because of the tight query budget and limited training samples. To improve the query efficiency, we use another signal from the oracle. The oracle $\mathcal{O}$ returns not only an image but also the tags of the returned image. We assume that these tags have similar average rewards and add these tags into the training dataset. Specifically, when we query tag $t \in \mathcal{T}$, we use $\{(s, f(\mathcal{O}(t).\texttt{image})) \mid s \in \mathcal{O}(t).\texttt{tags}\}$ as training data, whereas \textsc{Tiara}-S uses only $\{(t, f(\mathcal{O}(t).\texttt{image}))\}$. As we will see in the experiment, this technique significantly improves query efficiency and performance. Algorithm \ref{algo} shows the pseudo-code of \textsc{Tiara}.

\vspace{0.1in}
\noindent \textbf{Time Complexity.} The bottleneck of the computation is in Line 7. The inverse matrix $\boldA^{-1}$ can be efficiently computed in an iterative manner using the Sherman–Morrison formula, i.e.,
\[ (\boldA + \boldv_t \boldv_t^\top)^{-1} = \boldA^{-1} - \frac{\boldA^{-1} \boldv_t \boldv_t^\top \boldA^{-1}}{1 + \boldv_t^\top \boldA^{-1} \boldv_t}. \]
This technique reduces the cubic dependence on the number of dimensions to the quadratic dependence. Therefore, the computation is in Line 7 takes $O(|\mathcal{T}_\text{known}| d^2)$ time. Let $T_\text{max}$ be the maximum number of tags of an image. The value of $T_\text{max}$ is typically $10$ to $100$. Because $|\mathcal{T}_\text{known}|$ increases by at most $T_\text{max}$ in an iteration, the total time complexity is $O((|\mathcal{T}_\text{ini}| + B T_\text{max}) B d^2)$ in the worst case. In our problem setting, $B$ is small (e.g., in the hundreds) because of the tight API limitation, and $|\mathcal{T}_\text{ini}| \approx 100$ and $d \approx 300$ are also small during the experiments. Therefore, the oracle calls in Line 9 become a bottleneck in the wall-clock time because it requires communication to the Internet. As the oracle calls are common in all methods, \textsc{Tiara} is sufficiently efficient. When the black-box function is complex, its evaluation can be a computational bottleneck as well. \textsc{Tiara} evaluates $f$ as few as $B$ times, which is the same as with other baseline methods. In addition, when the efficiency is insufficient, we can speed up \textsc{Tiara} through a lazy variance update \cite{desautels2012parallelizing} and by applying sub-sampling heuristics.\footnote{We do not adopt such techniques in the experiments.}

\vspace{0.1in}
\noindent \textbf{Discussion on graph-feedback bandits.} There are several multi-armed bandit algorithms with a so-called feedback graph setting \cite{mannor2011from, alon2013from, kocak2014efficient}. This is an intermediate setting between the bandit and full feedback. Specifically, it assumes that there is an underlying graph, where a node represents an arm, and when we choose arm $t$, we can also observe the rewards of the neighboring arms. The underlying graph can be directed and time-varying. Therefore, if we define the neighbor of $t$ as $\mathcal{O}(t).\texttt{tags}$, our problem is seen as a variant of the feedback graph setting. However, we do not employ the feedback graph framework for the following reasons: First, existing graph feedback bandit algorithms require the number of neighbors of each neighbor \cite{mannor2011from, alon2013from, kocak2014efficient}, i.e., $|f(\mathcal{O}(s).\texttt{tags})|, \forall s \in f(\mathcal{O}(t).\texttt{tags})$ in our setting. We need additional API queries to compute these values. Such additional queries are prohibitive because the API limit is tight under our setting. Second, graph feedback bandit algorithms assume that the reward feedback of neighboring arms is an unbiased estimate of the true rewards \cite{mannor2011from, alon2013from, kocak2014efficient}. However, $f(\mathcal{O}(t).\texttt{image})$ is not unbiased for arm $s \in \mathcal{O}(t).\texttt{tags}$. Therefore, we cannot enjoy the theoretical guarantees of the graph feedback bandits. Third, the improvement obtained by the feedback graph framework is on the order of $O(\sqrt{\alpha/n})$ \cite{kocak2014efficient}, where $n$ is the number of arms, and $\alpha$ is the size of the maximum independent set of the feedback graph. Under our setting, the sizes of the feedback graphs and $\alpha$ are still large in practice. How to leverage the existing graph-feedback bandit algorithms for our setting is an interesting area of future study.

\vspace{0.1in}
\noindent \textbf{Discussion on the reward estimation model.} \textsc{Tiara} uses LinUCB and a linear model for estimating the reward from the feature vector. In general, we can use any model for prediction. We use a linear model owing to the following reasons: First, complex models are difficult to train with a tight sample budget. Second, LinUCB empirically performs quite well under various tasks \cite{li2010contextual, krishnamurthy2016contextual} regardless of its suboptimal theoretical regret. Third, we use the average word embeddings as the feature vector of an arm. It has been confirmed that word embeddings are representative and that the weighted average of the word embeddings with a cosine similarity or liner models produce superior performances in many NLP tasks \cite{arora2017simple, shen2018baseline} and sometimes outperform even neural network-based methods.

\subsection{Visualization and Interpretation}

Although not our original goal, \textsc{Tiara} provides an interpretation of the black-box function $f$ as a byproduct. 

Deep neural networks suffer from interpretability issues for reliable decision making. There have been many interpretation methods developed for deep neural networks \cite{simonyan2014deep, sundararajan2017axiomatic}. We consider model-level interpretability \cite{simonyan2014deep, nguyen2016synthesizing, yuan2020xgnn}, i.e., interpreting what function each model represents. It is unclear what function each model represents by simply looking at the model parameters of deep models. Even if the model is prepared by the user, deep models sometimes behave in unexpected ways \cite{goodfellow2015explaining, madry2018towards}. Input instances that produce high values can be regarded as representations of the model \cite{simonyan2014deep, nguyen2016synthesizing, yuan2020xgnn}. \textsc{Tiara} can retrieve such images from external image databases. Compared to methods that use a fixed dataset, the search space of \textsc{Tiara} is extremely large. Therefore, there are more chances that relevant images will be found. In addition, tag scores of \textsc{Tiara} also provide another interpretation of the black-box function through words. We show that these tags are also beneficial for interpretability by visualizing word clouds in the experiments. The tag-based interpretability is beneficial for further exploration as well. When the result is unsatisfactory, the user can continue manual exploration from the tags with high scores.

\section{Experiments}

\begin{figure*}[t]
\centering
\includegraphics[width=0.95\hsize]{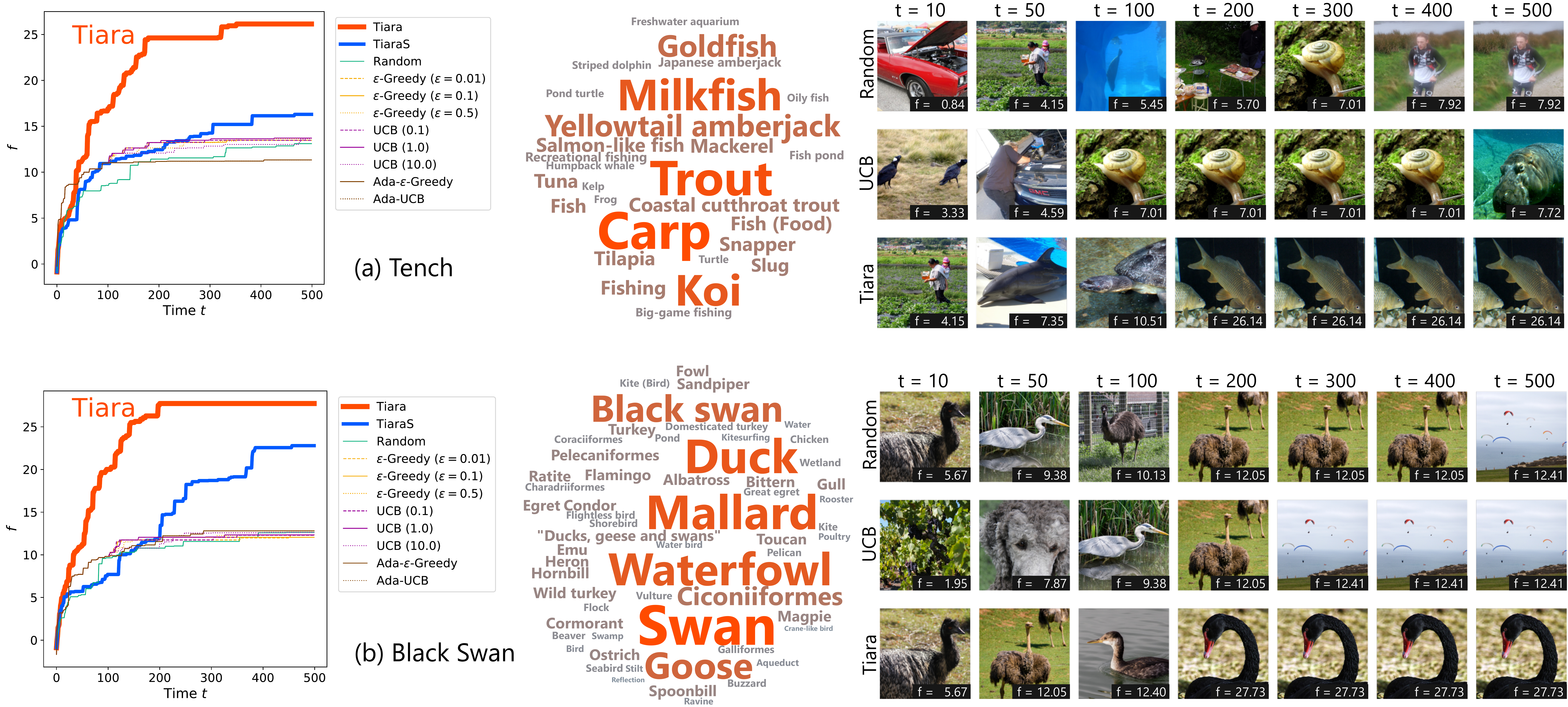}
\caption{Open Image Environment. (Top) Tench, (Bottom) Black Swan. (Left) Learning curves averaged over $10$ independent runs. (Mid) Word clouds that represent the scores computed by \textsc{Tiara} at the end of the last iteration. These visualizations provide an interprtability of the black-box function. (Right) Best images found at each iteration.}
\label{fig: openimage}
\vspace{-0.1in}
\end{figure*}

\begin{table*}[t]
    \centering
    \caption{Open Image Environment. Each score is the highest $f(x)$ found by an algorithm. The scores are averaged over $10$ independent runs, and the standard deviations are also reported. The highest score is highlighted by bold in each column.}
    \scalebox{0.7}{
\begin{tabular}{lccccccccccc} \toprule
Method $\backslash$ Class & Tench & Black Swan & Tibetan Terrier & Tiger Beetle & Academic Gown & Cliff Dwelling & Hook & Paper Towel & Slot Machine & Water Tower & Average \\ \midrule
Random & 13.11 $\pm$ 5.95 & 12.62 $\pm$ 4.83 & 13.16 $\pm$ 1.92 & 8.21 $\pm$ 1.85 & 12.63 $\pm$ 4.88 & 12.63 $\pm$ 5.31 & 11.58 $\pm$ 1.51 & 10.40 $\pm$ 1.01 & 9.29 $\pm$ 1.68 & 12.09 $\pm$ 1.64 & 11.57 $\pm$ 0.89 \\ 
$\epsilon$-Greedy ($\epsilon = 0.01$) & 13.48 $\pm$ 6.27 & 12.30 $\pm$ 4.94 & 13.76 $\pm$ 2.16 & 7.47 $\pm$ 2.53 & 14.83 $\pm$ 5.80 & 13.32 $\pm$ 6.07 & 11.17 $\pm$ 1.85 & 9.79 $\pm$ 0.84 & 8.55 $\pm$ 1.02 & 13.25 $\pm$ 1.41 & 11.79 $\pm$ 0.99 \\
$\epsilon$-Greedy ($\epsilon = 0.1$) & 13.48 $\pm$ 6.27 & 12.06 $\pm$ 4.99 & 13.76 $\pm$ 2.16 & 8.54 $\pm$ 3.19 & 14.83 $\pm$ 5.80 & 13.06 $\pm$ 6.26 & 11.17 $\pm$ 1.85 & 9.66 $\pm$ 0.83 & 8.84 $\pm$ 0.88 & 13.25 $\pm$ 1.41 & 11.86 $\pm$ 1.08 \\
$\epsilon$-Greedy ($\epsilon = 0.5$) & 13.64 $\pm$ 6.07 & 12.30 $\pm$ 4.96 & 13.76 $\pm$ 2.16 & 9.88 $\pm$ 3.39 & 13.10 $\pm$ 4.77 & 13.36 $\pm$ 6.02 & 11.64 $\pm$ 1.96 & 9.74 $\pm$ 0.95 & 8.42 $\pm$ 1.16 & 13.06 $\pm$ 1.63 & 11.89 $\pm$ 1.07 \\
UCB ($\alpha = 0.1$) & 13.48 $\pm$ 6.27 & 12.30 $\pm$ 4.94 & 13.76 $\pm$ 2.16 & 7.47 $\pm$ 2.53 & 14.83 $\pm$ 5.80 & 13.32 $\pm$ 6.07 & 11.17 $\pm$ 1.85 & 9.79 $\pm$ 0.84 & 8.55 $\pm$ 1.02 & 13.25 $\pm$ 1.41 & 11.79 $\pm$ 0.99 \\
UCB ($\alpha = 1.0$) & 13.71 $\pm$ 5.99 & 12.33 $\pm$ 4.94 & 13.76 $\pm$ 2.16 & 9.27 $\pm$ 3.73 & 14.83 $\pm$ 5.80 & 13.32 $\pm$ 6.07 & 11.17 $\pm$ 1.85 & 10.14 $\pm$ 0.98 & 8.42 $\pm$ 1.17 & 13.25 $\pm$ 1.41 & 12.02 $\pm$ 1.14 \\
UCB ($\alpha = 10.0$) & 13.68 $\pm$ 5.98 & 12.62 $\pm$ 4.83 & 13.76 $\pm$ 2.16 & 8.21 $\pm$ 1.85 & 13.05 $\pm$ 4.58 & 13.36 $\pm$ 6.02 & 11.61 $\pm$ 2.00 & 9.77 $\pm$ 1.25 & 8.70 $\pm$ 1.50 & 13.25 $\pm$ 1.41 & 11.80 $\pm$ 0.87 \\
Ada-$\epsilon$-Greedy ($\epsilon = 0.1$) & 11.34 $\pm$ 5.84 & 12.81 $\pm$ 4.97 & 13.28 $\pm$ 1.97 & 8.70 $\pm$ 1.36 & 15.62 $\pm$ 3.38 & 10.74 $\pm$ 2.13 & 11.09 $\pm$ 1.72 & 9.93 $\pm$ 0.81 & 8.43 $\pm$ 0.88 & 10.25 $\pm$ 1.34 & 11.22 $\pm$ 1.08 \\
Ada-UCB ($\alpha = 1.0$) & 11.34 $\pm$ 5.84 & 12.81 $\pm$ 4.97 & 13.28 $\pm$ 1.97 & 8.70 $\pm$ 1.36 & 15.62 $\pm$ 3.38 & 10.74 $\pm$ 2.13 & 11.09 $\pm$ 1.72 & 9.93 $\pm$ 0.81 & 8.43 $\pm$ 0.88 & 10.25 $\pm$ 1.34 & 11.22 $\pm$ 1.08 \\
TiaraS & 16.30 $\pm$ 8.05 & 22.79 $\pm$ 7.73 & 12.67 $\pm$ 4.23 & 15.13 $\pm$ 2.88 & 25.40 $\pm$ 1.11 & \textbf{22.21 $\pm$ 0.00} & 12.21 $\pm$ 1.65 & 9.73 $\pm$ 1.69 & 12.60 $\pm$ 2.56 & 14.81 $\pm$ 1.08 & 16.39 $\pm$ 1.21 \\
Tiara & \textbf{26.14 $\pm$ 0.00} & \textbf{27.73 $\pm$ 0.00} & \textbf{15.92 $\pm$ 0.00} & \textbf{16.91 $\pm$ 0.00} & \textbf{25.77 $\pm$ 0.00} & \textbf{22.21 $\pm$ 0.00} & \textbf{14.06 $\pm$ 0.16} & \textbf{22.13 $\pm$ 0.00} & \textbf{16.77 $\pm$ 0.30} & \textbf{15.69 $\pm$ 0.00} & \textbf{20.33 $\pm$ 0.03} \\ \bottomrule
    \end{tabular}
    }
    \label{tab: openimage}
\end{table*}

We investigate the performance of \textsc{Tiara} using various real-world datasets.

\subsection{Experimental Setups}

\subsubsection{Baselines}

We use the following baselines.

\begin{itemize}
    \item \textbf{Random} queries random known tags.
    \item \textbf{$\varepsilon$-Greedy} is a bandit algorithm. This algorithm chooses the best tag with the highest empirical mean reward with a probability of $1 - \varepsilon$ and chooses a random tag with probability $\varepsilon$. The candidate pool is set to $\mathcal{T}_\text{ini}$.
    \item \textbf{UCB} is a bandit algorithm. The score of tag $t$ is the empirical mean reward plus $\alpha \sqrt{1/n_t}$, where $n_t$ is the number of observations from tag $t$, and $\alpha$ is a hyperparameter. UCB chooses the tag with the highest score. The candidate pool is set to $\mathcal{T}_\text{ini}$.
    \item \textbf{Ada-$\varepsilon$-Greedy} is a variant of $\varepsilon$-Greedy. This algorithm inserts new tags to the candidate pool $\mathcal{T}_\text{known}$ when new tags are found and chooses a tag from $\mathcal{T}_\text{known}$.
    \item \textbf{AdaUCB} is a variant of UCB. This algorithm inserts new tags to the candidate pool $\mathcal{T}_\text{known}$ when new tags are found and chooses a tag from $\mathcal{T}_\text{known}$.
    \item \textbf{\textsc{Tiara}-S} is a variant of \textsc{Tiara} that uses only the query tag for training.
\end{itemize}

\subsubsection{Hyperparameters}

We use $\lambda = 1$ and $\alpha = 0.01$ for \textsc{Tiara} and \textsc{Tiara}-S across all settings without further tuning. We will show that \textsc{Tiara} is insensitive to the choice of these hyperparameters over orders of magnitude in Section \ref{exp: sensitivity}. We use $300$-dimensional GloVe\footnote{\url{https://nlp.stanford.edu/projects/glove/}} trained using six billion Wikipedia 2014 + Gigaword 5 tokens for the word embeddings.

We report the performance with various hyperparameters for the baseline methods. Note that hyperparameter tuning is prohibitive in practice because of the tight API limit. If we apply hyperparameter tuning, we should use these query budgets for the main task instead. Therefore, this setting is slightly advantageous for the baseline methods.

\subsection{Open Images dataset}

\begin{figure*}[t]
\centering
\includegraphics[width=\hsize]{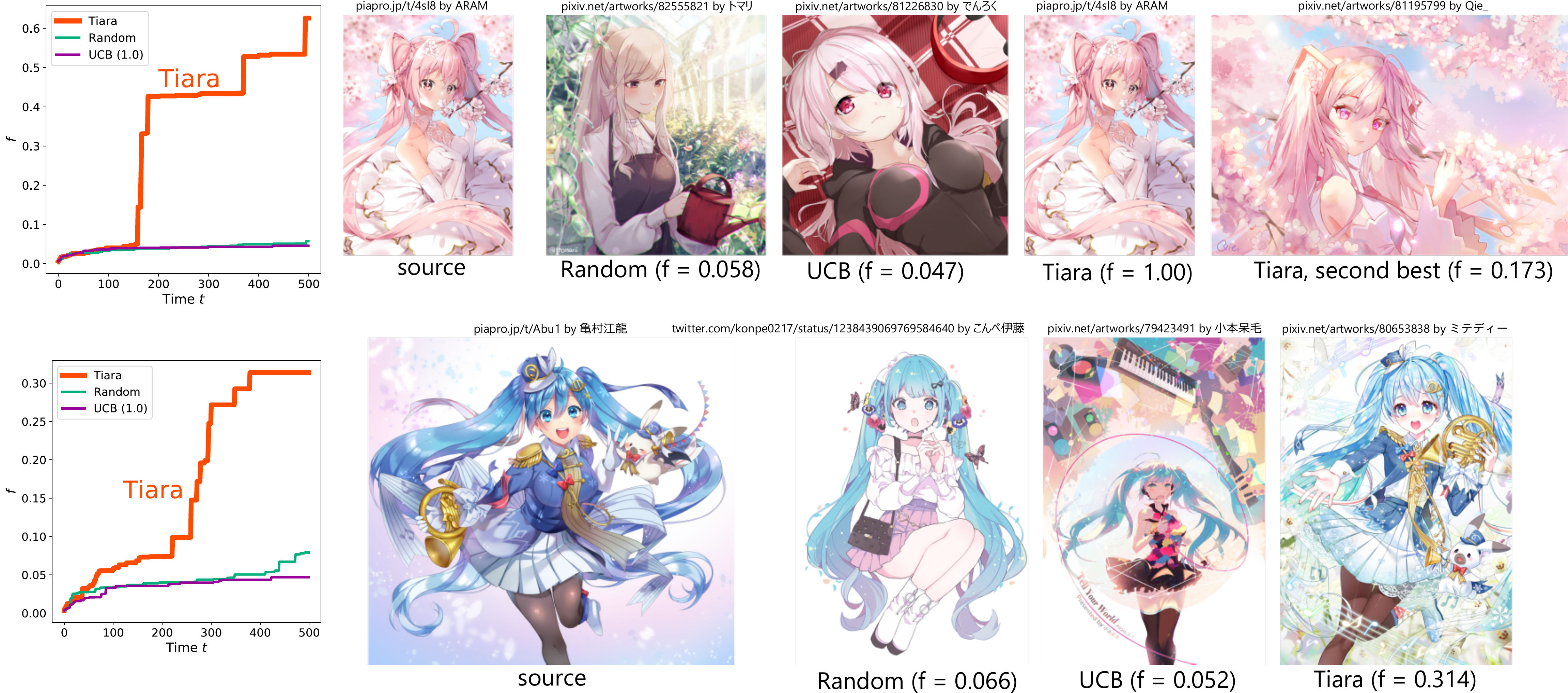}
\vspace{-0.15in}
\caption{Safebooru Environment.\protect\footnotemark (Left) Learning curves averaged over $10$ independent runs. (Middle) Source images. (Right) Best Images found. In the first row, we show the second best image for \textsc{Tiara} because \textsc{Tiara} found the exact source image.}
\label{fig: miku}
\vspace{-0.1in}
\end{figure*}

We use the Open Images Dataset V6 \cite{kuznetsova2020open} to construct the environment for the first testbed. Each image in this dataset has multi-class annotations, such as ``cat,'' ``Aegean cat,'' and ``Pumpkin pie.'' An image has $8.8$ classes on average. We use these classes as tags. We utilize ResNet18 \cite{he2016deep} trained with ImageNet for the black-box functions. Specifically, for each class $c$ of ImageNet, we use the pre-softmax logit for $c$ as the back-box function. Among $1000$ classes of ImageNet, we use $10$ classes $c = 1, 101, \cdots, 901$, i.e., Tench, Black Swan, Tibetan Terrier, Tiger Beetle, Academic Gown, Cliff Dwelling, Hook, Paper Towel, Slot Machine, and Water Tower. Note that the retrieval algorithms do not know these class names, i.e., the function is a black-box. We use these class names for only evaluations and interpretations of the results. Note also that the set of tags (i.e., classes of the Open Image Dataset) differs from the set of classes of ImageNet.

\footnotetext{The source images are used under the attribution, non-commercial, and no-derivative piapro license. The other images are used with artist permission.}

We subsample $10{,}000$ test images from the Open Images Dataset and construct an environment. When we query tag (class) $t$, this environment returns a random image with class $t$ and the set of classes this image belongs to. We choose $100$ random tags as the initial known tags $\mathcal{T}_\text{ini}$ and set the budget to $B = 500$.

We run ten trials with different seeds. Table \ref{tab: openimage} reports the means and standard deviations of the best $f$ for each method within $B$ queries. The last column reports the average of ten classes. These results show that \textsc{Tiara} performs the best under all settings, and \textsc{Tiara}-S performs second best on average.

The middle panels in Figures \ref{fig: openimage} show the word clouds \footnote{\url{https://github.com/amueller/word_cloud}} generated by \textsc{Tiara}. The size and color of a tag represent the score of the tag at the final iteration. These scores provide interpretations for the black-box functions. For example, in the case of tench, fish-related tags, such as Trout, Carp, and Milkfish, have high scores. In the black swan case, bird-related tags, such as Swan, Waterfowl, and Duck, have high scores. We stress again that \textsc{Tiara} does not use the ground truth class name but instead treats $f$ as a black-box function. Even when the ground truth class name is unavailable to us, the word cloud generated by \textsc{Tiara} and the retrieved images indicate that $f$ is fish-related in the first example and bird-related in the second example with significant interpretability.

\begin{figure*}[t]
\centering
\includegraphics[width=0.95\hsize]{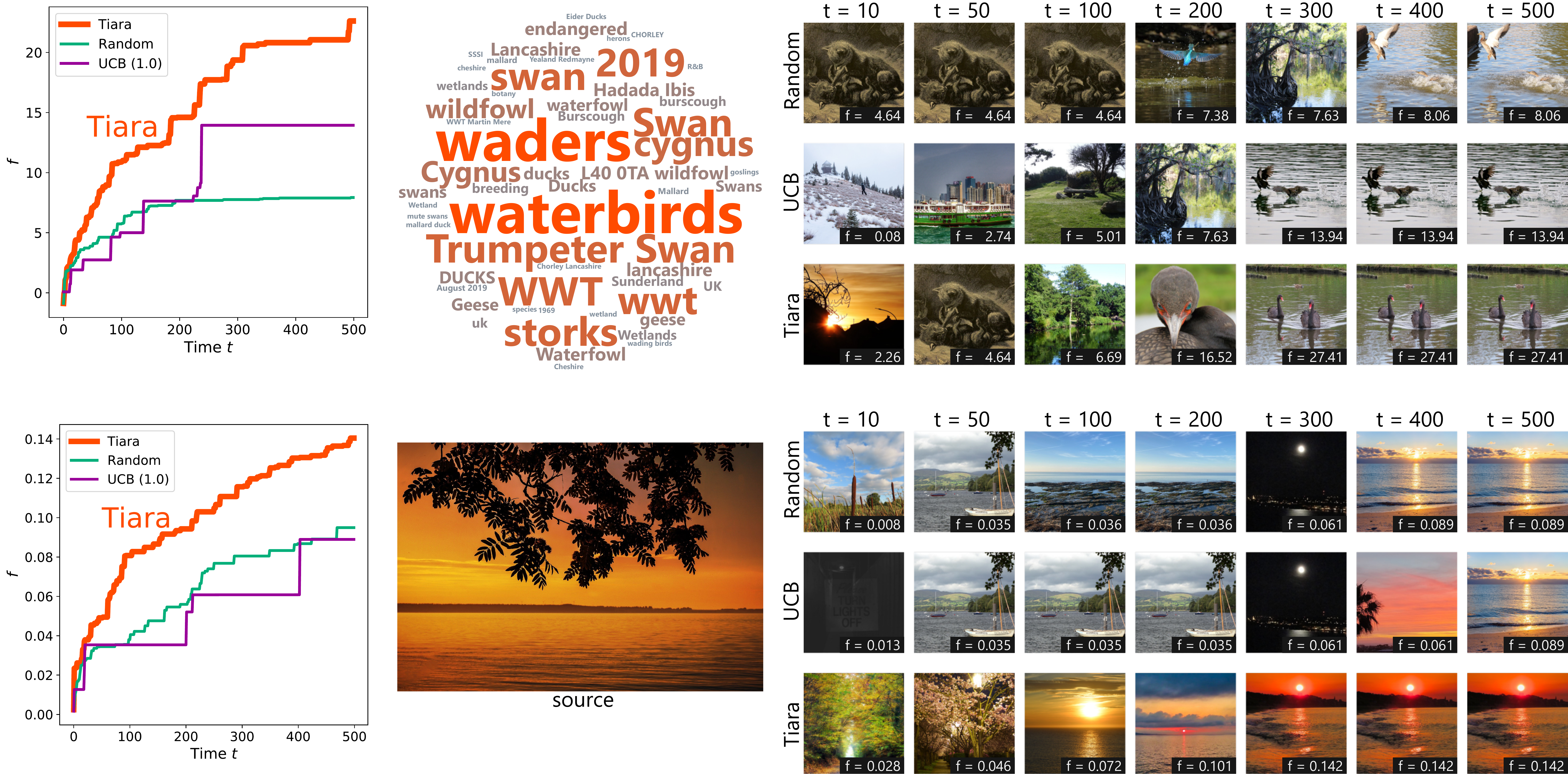}
\vspace{-0.1in}
\caption{Flickr Environment. (Top) Black swan. (Bottom) Similar image retrieval. (Left) Learning curves averaged over ten independent runs. (Top middle) Word cloud generated by \textsc{Tiara}. (Bottom middle) Source image. (Right) Best images found duing each iteration.}
\label{fig: flickr}
\vspace{-0.1in}
\end{figure*}

\subsection{Safebooru Environment}

The aim of this section is to confirm that \textsc{Tiara} is effective even in a completely different domain. We use Safebooru as a testbed. We use a dump retrieved on June 7, 2019 \footnote{\url{https://www.kaggle.com/alamson/safebooru}}. Each image has $34$ tags on average, such as ``smile,'' ``long hair,'' and ``blonde hair.'' We use illustration2vec \cite{saito2015illustration} to construct the black-box functions.

We use the latest $100{,}000$ images from this dataset and construct an environment. As a result of the broken links, this environment contains $81{,}517$ images in total. When we query tag $t$, this environment returns a random image with tag $t$ and the set of tags this image belongs to. We choose $100$ random tags as the initial known tags $\mathcal{T}_\text{ini}$ and set the budget to $B = 500$.

We conduct semantically similar image search experiments. We emphasize that although this environment does not provide an official image-based search, our algorithm enables such an image-based search. Given a source image $s$, we compute $4096$-dimensional embedding $\boldv_s \in \mathbb{R}^{4096}$ using the pre-trained illustration2vec model \footnote{\url{https://github.com/rezoo/illustration2vec}}. We use the Gaussian kernel between the embeddings of an input image $x$ and the source image as the black-box function, i.e., $f(x) = \exp(-\|\boldv_s - \boldv_x\|^2/\sigma^2)$, where $\boldv_x \in \mathbb{R}^{4096}$ is the embedding of input $x$ computed using the illustration2vec model, and $\sigma$ is the bandwidth of the Gaussian kernel. We set $\sigma = 100$ during this experiment. The more semantically similar the input image is to the source, the higher the value taken by this function.

We use two illustrations shown in Figure \ref{fig: miku} (middle) as the source images. The right panels in Figures \ref{fig: miku} show images retrieved by the methods. The first source image is in the image database, and \textsc{Tiara} succeeds in retrieving the same image from the database. We show the second best image retrieved by \textsc{Tiara} in this case. Even the second best image is semantically similar to the source image, i.e., it depicts the same character with cherry blossom motifs, and has a higher objective value than the images retrieved by the baseline methods. For the second case, the source image is not in the database. Although not exactly the same, \textsc{Tiara} succeeds in retrieving a semantically similar image with the same characters and motifs, i.e., horn, rabbit, and costume.

These results show the flexibility of our framework such that \textsc{Tiara} can be applied to not only photo-like image databases but also illustration-like image databases using appropriate black-box functions. e.g., the illustration2vec model.

\subsection{Flickr Environment}

The aim of this section is to confirm that \textsc{Tiara} is effective and readily applicable to real-world environments. We use the online Flickr environment in operation as a testbed. We also use ResNet18 trained with ImageNet to construct the black-box functions. We implement oracle $\mathcal{O}$ by combining \texttt{flickr.photos.search} and \texttt{flickr.tags.getListPhoto} APIs. We use the \texttt{license='9,10'} option such that it returns only public domain or CC0 images. We also set the budget to $B = 500$. Because Flickr contains as many as \emph{ten billion} images, it is challenging to find relevant images within $B = 500$ queries. Each image has $15$ tags on average, including ``summer,'' ``water,'' and ``sea.'' 

We conduct two experiments in this environment. First, we conduct the same experiment as in the open image environment. The top row of Figure \ref{fig: flickr} shows the result for the black swan class. Compared to Figure \ref{fig: openimage}, \textsc{Tiara} in this environment learns more slowly than in the open image environment. We hypothesis that this is because the Flickr environment contains noisy tags annotated by users, which are occasionally described in foreign languages, whereas the open image environment contains only clean tags that indicate solid categories judged by annotators. Nevertheless, \textsc{Tiara} succeeds in retrieving black swan images in several hundred queries.

Second, we conduct similar image retrieval experiments as in the safebooru environment. We use the output of the penultimate layer of the pre-trained ResNet18 for the image embedding. We use the Gaussian kernel between the embeddings of an input image $x$ and the source image $s$ as the black-box function, i.e., $f(x) = \exp(-\|\boldv_s - \boldv_x\|^2/\sigma^2)$, where $\boldv_x \in \mathbb{R}^{512}$ is the embedding of input $x$ computed using ResNet18, and $\sigma$ is the bandwidth of the Gaussian kernel. We set $\sigma = 10$ in this experiment. Figure \ref{fig: flickr} (bottom, center) shows the source image. As the right panels show, \textsc{Tiara} succeeds in retrieving semantically similar images depicting a sunrise over the sea.

\begin{figure*}[t]
\centering
\includegraphics[width=0.33\hsize]{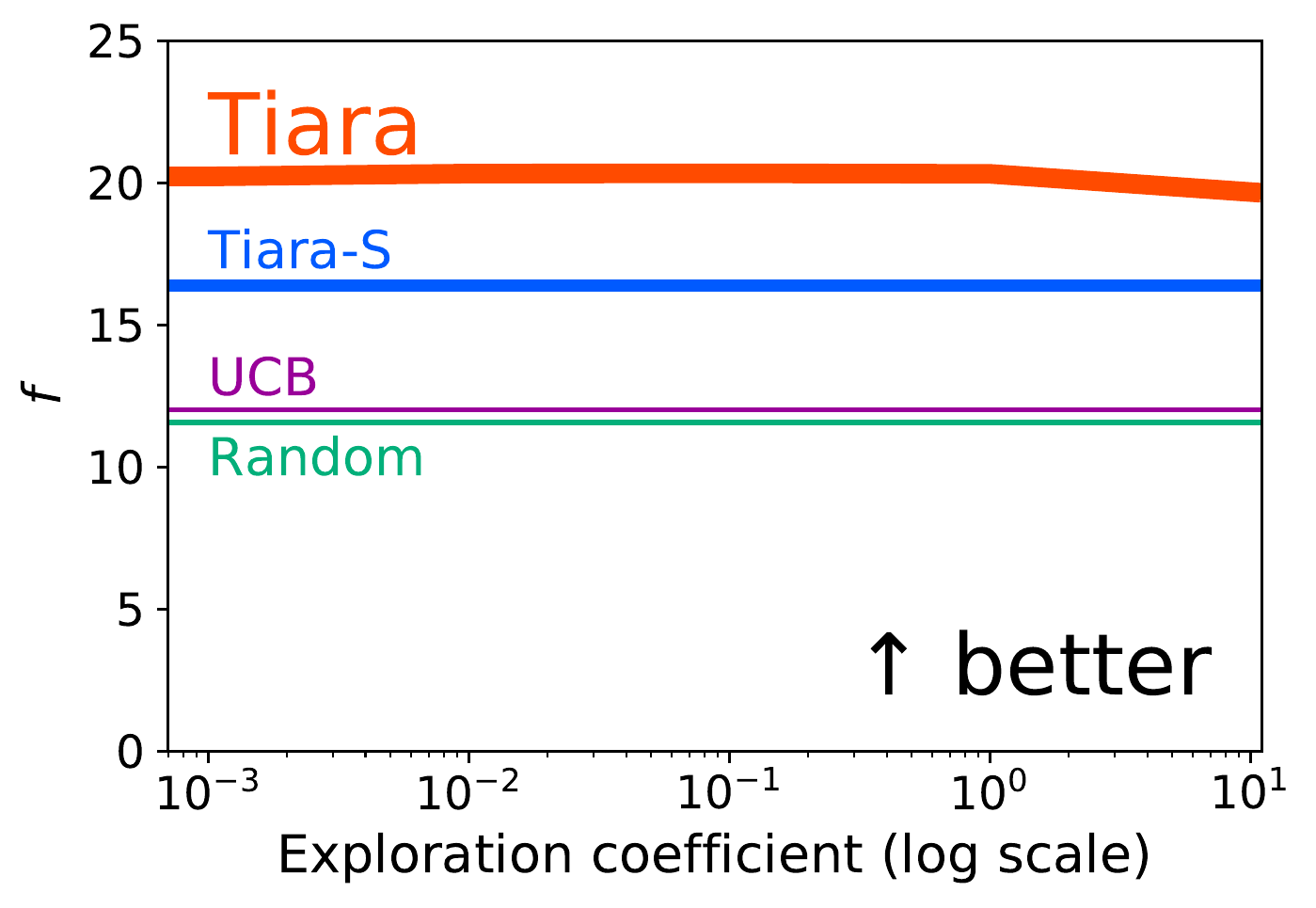}
\includegraphics[width=0.33\hsize]{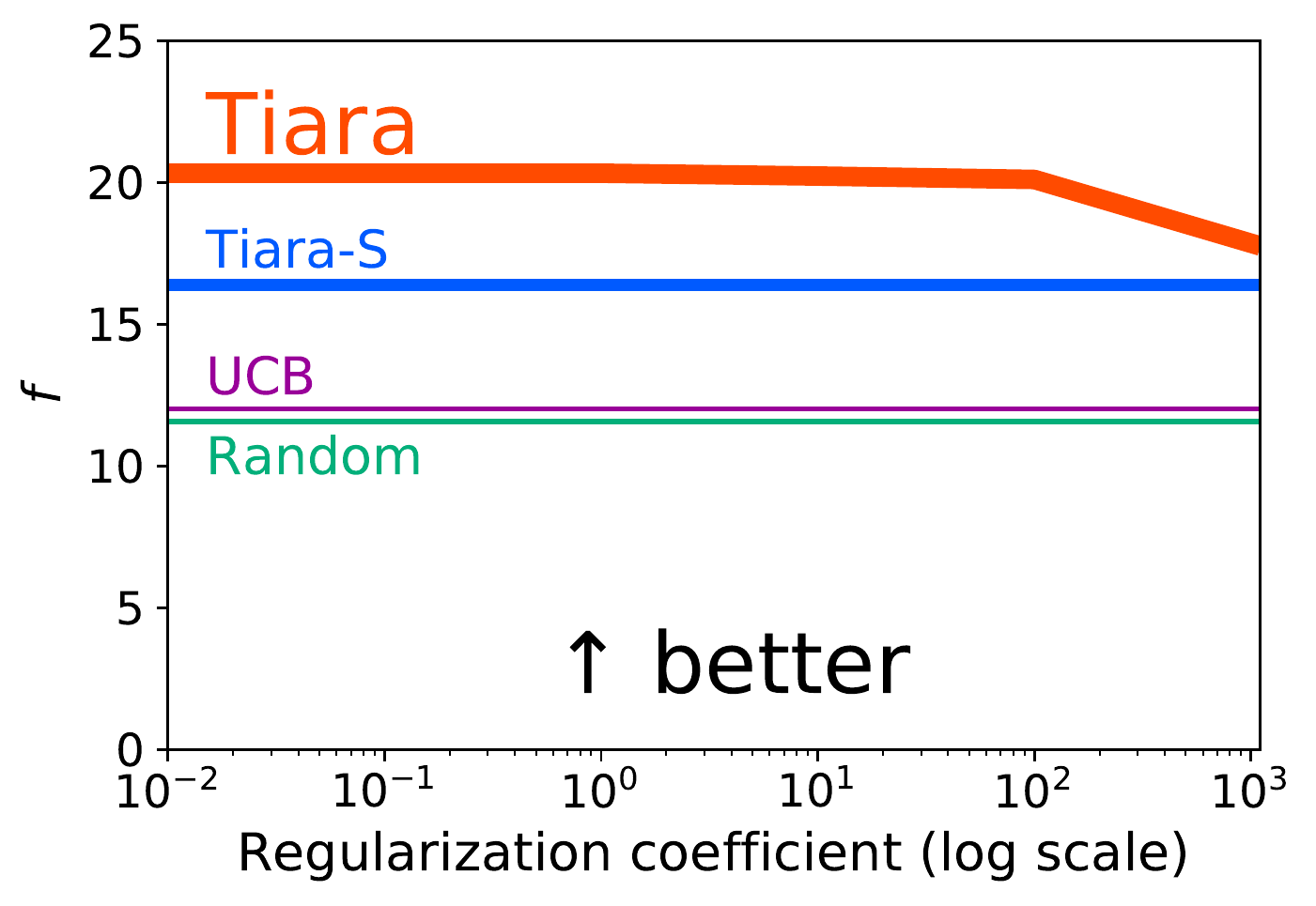}
\includegraphics[width=0.33\hsize]{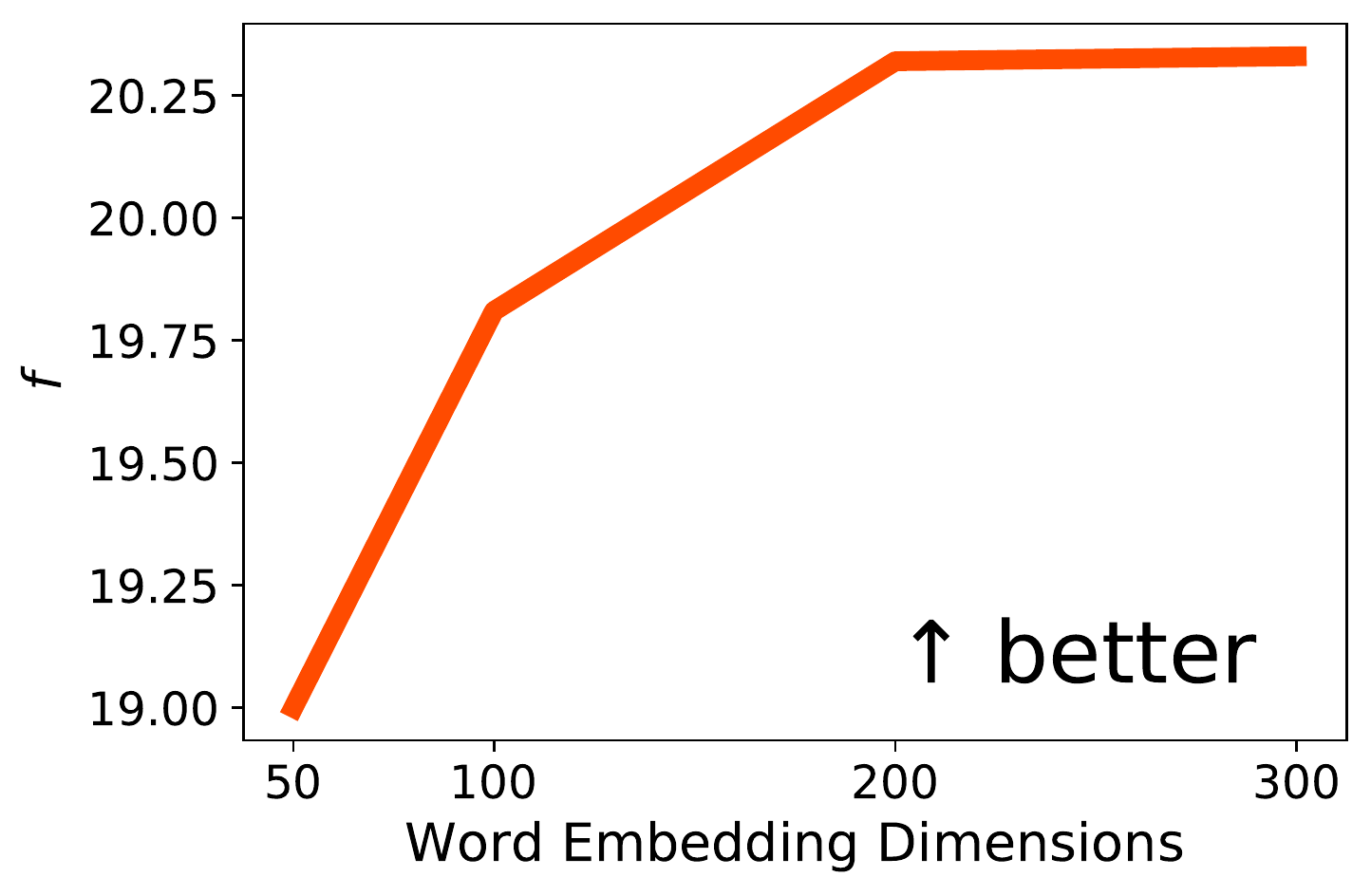}
\caption{Hyperparameter sensitivity. As the default settings, we set the exploration coefficient as $\alpha = 0.01$, regularization coefficient as $\lambda = 1.0$, and word embedding as $300$-dimensional Glove and change one configuration in each panel. These results show that \textsc{Tiara} is stable with respect to the hyperparameter choices across several orders of magnitude.}
\label{fig: hyperparameter}
\end{figure*}

\subsection{Hyperparameter Sensitivity} \label{exp: sensitivity}

We investigate the hyperparameter sensitivity of \textsc{Tiara} in this section. \textsc{Tiara} has two hyperparameters, i.e., exploration coefficient $\alpha$ and regularization coefficient $\lambda$. \textsc{Tiara} also has a choice of word embeddings. It is crucial to ensure stability with respect to the hyperparameter choices because tuning the hyperparameters in a new environment is difficult with tight API budgets. The left panel of Figure \ref{fig: hyperparameter} shows the average performance of \textsc{Tiara} in the Open Image Dataset environment with various values of $\alpha$ maintaining $\lambda = 1.0$ (default value). The $x$-axis is plotted on a log scale. We show the performances of \textsc{Tiara}-S, UCB, and Random for reference. In this plot, we do not tune the hyperparameters of \textsc{Tiara}-S accordingly to maintain the conciseness of the plot. This plot shows that \textsc{Tiara} is stable with respect to $\alpha$ over several orders of magnitude. The middle panel of Figure \ref{fig: hyperparameter} shows the average performance with various values of $\lambda$ maintaining $\alpha = 0.01$ (default value). The $x$-axis is plotted on a log scale. The result shows that \textsc{Tiara} is stable with respect to $\lambda$ over several orders of magnitude. The right panel of Figure \ref{fig: hyperparameter} shows the average performance with various dimensions of the GloVe word embedding maintaining $\alpha = 0.01$ and $\lambda = 1.0$ (default values). This result shows that \textsc{Tiara} performs better with higher dimensional embeddings, i.e., more expressive embeddings. It also indicates that \textsc{Tiara} indeed utilizes the word embedding geometry. We recommend using high-dimensional and strong word embeddings for \textsc{Tiara}.

\section{Related Work}

\subsection{Image Retrieval}

Image retrieval has been studied for decades. The main concern is how to retrieve relevant images \cite{babenko2015aggregating, kato2008can, leiva2011query} with efficiency \cite{lai2015simultaneous, liu2016deep}. Deep neural networks have been preferred for image retrieval in recent years because they can extract rich features including texture \cite{cimpoi2015deep}, style \cite{gatys2016image}, and semantics \cite{babenko2015aggregating}. Hash-based methods have also been extensively studied for their efficiency \cite{lai2015simultaneous, liu2016deep}. In addition, many extensions have been studied, such as multi-modal image search \cite{cao2016deep, kordan2018deep} and contextual retrieval \cite{xie2019improving}. We investigated the image retrieval problem from a different perspective. Specifically, we consider how an outsider can retrieve desirable images from an \emph{external} image database with as few queries as possible.

\subsection{Web Crawling}

Our problem setting can be seen as a crawling problem. Developing efficient web crawlers is a long-standing problem in the literature \cite{cho1998efficient, chakrabarti1999focused, diligenti2000focused, cho2003effective, castillo2005effective, pham2019bootstrapping}. In particular, focused crawling \cite{chakrabarti1999focused, mccallum2000automating, johnson2003evolving, baezayates2005crawling, guan2008guide} is relevant to our problem setting. Focused crawling aims to efficiently gather relevant pages by skipping irrelevant pages.

However, there are several differences between our study and existing focused crawling. First, focused crawling is used to search web pages by following WWW hyperlinks, whereas we search for \emph{images} from an \emph{external database} utilizing the tag search oracle. Thus, the existing crawling methods are not directly applicable to our setting. Second, we assume the API limit is extremely tight, e.g., $500$, whereas existing focused crawlers typically visit hundreds of thousands of pages. Third, existing focused crawlers focus on retrieving pages of a specific topic \cite{chakrabarti1999focused, mccallum2000automating}, popular pages \cite{baezayates2005crawling}, pages with structured data \cite{meusel2014focused}, or hidden web pages \cite{barbosa2007adaptive}. By contrast, deep convolutional neural networks in our setting realize rich vision applications, as we show in the experiments. These applications are qualitatively different from the existing focused crawlers and are valuable in their own right.

\subsection{Private Recommender Systems}

Private recommender systems \cite{sato2021private} aim to build a fair recommender system when the service provider does not offer a fair system. Our problem can also be seen as constructing a private recommender system when the black-box function is a recommendation score. There are several differences between our approach and private recommender systems. First, the existing methods, \textsc{PrivateRank} and \textsc{PrivateWalk}, assume the use of an item recommendation oracle, which is unavailable under our setting. Second, \textsc{Tiara} considers content-based retrieval, whereas \textsc{PrivateRank} and \textsc{PrivateWalk} mainly focus on a collaborative recommendation scenario. Third, our application is not limited to fairness, and we showed promising applications, including semantically similar image retrieval.

\section{Discussion and Limitations}

Although we use deep neural networks as a black-box function, our framework and the proposed method are not limited to deep neural networks. For example, we can use human judgment as the black-box function, i.e., when we evaluate $f(x)$, we actually ask a human viewer how much he/she likes image $x$. Because our proposed method requires several hundred evaluations for a single run, the current method is too inefficient for human-in-the-loop experiments. More efficient methods are important for developing such intriguing applications.

In this work, we have assumed that $f$ is a well-behaved function. It may be interesting to use buggy $f$ functions. For example, when we debug a deep neural network model $f$, applying \textsc{Tiara} to $f$ may reveal what $f$ has already learned and has not yet. We hypothesize that there are much room for further applications of \textsc{Tiara}.

We have assumed that the only way to access the image database is the tag search oracle $\mathcal{O}$. Although our method is general owing to this formulation, many other APIs are available in real applications, such as user-based searches, popularity ranking, and collaborative filtering recommendations. Utilizing richer information to enhance performance and query efficiency is important in practice. At the other extreme, dropping the assumptions of tag affinity and tag search APIs to make the method applicable to broader databases is also an intriguing direction.

Although we have focused on image retrieval problems, our formulation is not limited to such problems and can be applied to other domains such as music, document, and video retrieval problems. Exploring further applications of our framework is left as future work.

\section{Conclusion}

In this paper, we formulated the problem of optimal image retrieval with respect to a given black-box function from an external image database. This problem enables each user to retrieve their preferable images from the Internet, even if the image database does not provide such features. We combined a bandit formulation with pre-trained word embeddings and proposed an effective retrieval algorithm called \textsc{Tiara}. Finally, we confirmed the effectiveness of \textsc{Tiara} using three environments, including an online Flickr environment.

\begin{acks}
This work was supported by JSPS KAKENHI GrantNumber 21J22490. \begin{CJK}{UTF8}{ipxm}We are grateful to ARAM, トマリ, でんろく, Qie\_, 亀村江龍, こんぺ伊藤, 小本呆毛, and ミテディー for allowing the use of their wonderful illustrations.\end{CJK}
\end{acks}

\bibliographystyle{plainnat}
\bibliography{sample-base}










\end{document}